\documentclass[english,showpacs,twocolumn]{revtex4}
\usepackage[T1]{fontenc}
\usepackage[latin1]{inputenc}
\usepackage{babel}
\usepackage{array}
\usepackage{graphics}
\usepackage{subfigure}
\usepackage{amsmath}
\makeatother
\begin{document}
\title{A variant of Peres-Mermin proof for testing noncontextual realist models}
\author{Alok Kumar Pan\footnote{apan@bosemain.boseinst.ac.in}}

\affiliation{CAPSS, Department of Physics, Bose Institute, Sector-V, Salt Lake, Calcutta 700091, India}

%\author{Alok Kumar Pan\footnote{apan@bosemain.boseinst.ac.in}}

%\affiliation{CAPSS, Department of Physics, Bose Institute, Sector-V, Salt Lake, Calcutta 700091, India}
\begin{abstract}
For any state in four-dimensional system, the quantum violation of an inequality based on the Peres-Mermin proof for testing noncontextual realist models has experimentally been corroborated. In the Peres-Mermin proof, an array of nine holistic observables for two two-qubit system was used. We, in this letter, present a new symmetric set of observables for the same system which also provides a contradiction of quantum mechanics with noncontextual realist models in a state-independent way. The whole argument can also be cast in the form of a new inequality that can be empirically tested.
\end{abstract}
\maketitle
\section{Introduction}

By demonstrating an ingenious \emph{gadanken} experiment  Einstein, Podoloski and Rosen had remarked\cite{epr} that the quantum mechanical description of nature is inherently incomplete. This is mainly because within the framework of quantum mechanics(QM) the very occurrence of a definite outcome in an individual measurement can not be ensured. The realist hidden variable models are those which seek to provide a `complete specification' of the state of a quantum system so that the individual measured values of any dynamical variable are predetermined by the appropriate values of the hidden variables. Studies on this issue have resulted in spectacular discoveries about the constraints that need to be imposed on the realist models in order to be consistent with the experimentally reproducible results of QM.  Bell's theorem\cite{belllocal} provides such a constraint by demonstrating a violation of an inequality by QM which is otherwise satisfied by all local realistic theory.  Another constraint, known as Bell-Kochen-Specker(BKS) theorem\cite{bell,kochen}, demonstrates the inconsistency between QM and the noncontextual realist(NCR) models by showing a contradiction when assigining non-contextual definite values to certain set of quantum observables. This paper concerns the latter strand of study. 

Different versions of BKS theorem were given %the proof of the mathematical theorem showing an incompatibility between the formalism of QM and the deterministic NCR models were given by Bell \cite {bell}, Kochen and Specker \cite{kochen}, followed by others 
suggesting a variety of ingenious proofs, see, for example, \cite{ker,cabello18, mermin, peres, penrose,cabello2, simon,home,michler,hasegawa1,hasegawa2, hasegawa3,cabellosi,nature,liu,ams}. The original proof by
Kochen and Specker was demonstrated by using 117 different real vectors in three-dimensional space. Subsequently, simpler proofs have been given by Kernaghan and Peres\cite{ker} using 20 directions, and, later by Cabello\cite{cabello18} using 18 directions. A geometrically elegant proof of BKS theorem for a spin-3/2 particle was given by Penrose\cite{penrose} by considering spin component measurements along 20 directions. In a different line of study, using an entangled state of a pair of spin-1/2 particles, Peres\cite{peres} demonstrated an inconsistancy between QM and the NCR models that involves six observables. Motivated by this work, Mermin\cite{mermin} presents a state-independent proof (henceforth, Peres-Mermin proof) in four-dimensional space using an array of nine holistic observables. In contrst to the earlier proofs\cite{kochen,ker,cabello18,penrose} the Peres-Mermin proof is of special interest because of its mathematical simplicity. Later, Cabello and Garcia-Alcaine \cite{cabello2} gave an argument using a two-particle two-state system that enables a suitable joint measurement pertaining to a particular set of compatible propositions to discriminate between QM and a testable consequence of the NCR hypothesis. Although this led to some interesting work \cite{simon}, a ticklish point is that this type of \textit{non-statistical} argument in terms of the yes-no validity of propositions is contingent upon the relevant dynamical variables being measured with infinite precision - an issue that has been the subject of considerable discussions  \cite{meyer} as to what extent the \textit{finite precision} (in the sense of `imprecision' in actually what is being measured) measurements can enable an empirical discrimination between QM and the deterministic NCHV models.

In this issue, a testable Bell-type inequality\cite{home,michler} was derived by simply replacing the locality condition with noncontxtuality assumption that is valid for any NCR model. The quantum violation of this inequality by an finite ammount(even if the actual measurements are inevitably imprecise) were acheived by Michler \emph{et al.}\cite{michler} for an intraparticle entangled state between path and polarization degrees of freedom of single photons, and by Hasegawa \emph{et. al.}\cite{hasegawa1} for the path and the spin degrees of freedom of single neutrons. Thereafter, the experimental investigation along this line was enriched by more studies \cite{hasegawa2,hasegawa3}. 

Relatively recently, a crucial developement in this issue has been made by Cabello\cite{cabellosi} by casting the Peres-Mermin proof in the form of a testable inequality involving the statistically measurable quantities - this inequality being violated by QM by a finite measurable amount for an \emph{arbitrary} two-qubit state. Subsequently, the quantum violation of this inequality has been experimentally corroborated using a pair of trapped ions\cite{nature} and using the polarization and the path degrees of freedom of  a single photon\cite{liu,ams}. 

In this paper, we provide a variant of the Peres-Mermin proof by replacing four out of nine holistic observables that has been used by Mermin\cite{mermin} by four new observables that results in deriving a new inequality which also provides a proof against the NCR models. Before proceeding further, let us first recapitulate the notion of `noncontextuality' as applied to any realist model. 

Let $\widehat\alpha$ and $\widehat\beta$ are two mutually commuting observables. Now, in any given realist hidden variable model, let $v(\alpha)$ be the individual measured values  of $\widehat\alpha$, as specified by a hidden variables $\lambda$,  and let $v(\beta)$ and $v(\alpha\beta)$ are the individual measured values of the observables $\widehat\beta$ and $\widehat{\alpha\beta}$ respectively, that are also predetermined by the same hidden variables. Then the notion of the property of the `noncontextuality' of the realist models is characterized by the following feature
\begin{equation}
v(\alpha\beta)= v(\alpha)v(\beta)
\end{equation}
This feature is known as `product rule'(for an elegant discussion, see, Ref.\cite{mermin}) that is assumed to hold good in a NCR model independent of the context for measuring $\widehat{\alpha\beta}$, and also independent of the way $\widehat{\alpha}$ and $\widehat{\beta}$ are separately measured. This rule has been invoked in the state-independent proof of BKS theorem by Mermin\cite{mermin} where he used an array of nine holistic observables.  

Here we use a new set of holistic observables for which the state-independent proof of BKS theorem follows upon exploiting this product rule. This proof can be cast in the form of a testable inequality which is again violated by QM. Before presenting our proof, for the sake of completeness we first reconsider Peres-Mermin proof. 

\section{The Peres-Mermin proof}
Let us consider an array of nine holistic observables of two two-level system that is suitably chosen jointly for the systems `1' and `2', each of which has the  eigenvalue $\pm1$. The array is given by 
\vskip -1.5cm
\begin{center}
\begin{equation}
\begin{array}{clrr}     
\sigma^{1}_{x}\otimes I^{2} & \hskip 0.7cm  I^{1}\otimes \sigma^{2}_{x} &\hskip 0.5cm  {\sigma^{1}_{x}\otimes\sigma^{2}_{x}} \\
\\
I^{1} \otimes \sigma^{2}_{y}  &\hskip 0.7cm  \sigma^{1}_{y}\otimes I^{2}  & \hskip 0.5cm  \sigma^{1}_{y}\otimes\sigma^{2}_{y}\\
\\
\sigma^{1}_{x}\otimes\sigma^{2}_{y} &\hskip 0.7cm  \sigma^{1}_{y}\otimes\sigma^{2}_{x}& \sigma^{1}_{z}\otimes\sigma^{2}_{z}
\end{array}
\end{equation}
\end{center}
Here, the Pauli operator $\sigma^{1}_{x}$ denotes the spin component along the $x$-axis of particle `1', and so on. Note that, the holistic observables in each row and each column  of the array given by Eq.(2) are mutually commuting and can then be simultaneously measured.

Now, for any quantum state in this four dimensional space $|\Psi\rangle$, is the eigenstate of the product of the three holistic observables in each row ($R_{i=1,2,3}$) and each column($C_{i=1,2,3}$), so that, 
\begin{subequations}
\begin{eqnarray}
\label{pm1}
&&\hskip -0.8cmR_1\left|\Psi\right\rangle= R_{2}\left|\Psi\right\rangle= R_3\left|\Psi\right\rangle =  C_1\left|\Psi\right\rangle= C_2\left|\Psi\right\rangle=|\Psi\rangle\\
\nonumber
\\
\label{pm2}
&&\hskip -0.8cmC_3\left|\Psi\right\rangle= -|\Psi\rangle
\end{eqnarray} 
\end{subequations}
where, $R_{1}=\left(\sigma^{1}_{x}\otimes I^{2}\right) .\left(I^{1} \otimes \sigma^{2}_{x} \right) . \left(\sigma^{1}_{x}\otimes\sigma^{2}_{x}\right)$ and similarly for the others.

In NCR models, it is assumed that the predetermined individual measured values as specified by hidden variable satisfy the same quantum mechanical eigenvalue relations given by Eqs.(\ref{pm1}) and (\ref{pm2}), then one can write the following relations 
\begin{subequations}
\begin{eqnarray}
&&\hskip -0.8cm v(R_1)= v(R_{2})= v(R_3)=  v(C_1)= v(C_2)=1\\
\nonumber
\\
&&\hskip -0.8cm v(C_3)= -1
\end{eqnarray} 
\end{subequations}
Note that, using the `product rule' that characterizes the NCR models, the individual measured values of the product of the three holistic observables(as example, $R_{1}$) can be written as the product of the values of those holistic observables, so that, for instance, $v(R_{1})$ can be written as 
\begin{eqnarray}
\label{prule1}
v(R_{1})=v\left(\sigma^{1}_{x}\otimes I^{2}\right) v \left(I^{1} \otimes \sigma^{2}_{x} \right) v \left(\sigma^{1}_{x}\otimes\sigma^{2}_{x}\right)
\end{eqnarray}

Applying the product rule for all the rows($v(R_{i=1,2,3})$ ) and columns($v(C_{i=1,2,3})$ ) it can be seen that the multiplication of all $v(R_{i=1,2,3})$ and $v(C_{i=1,2,3})$ yields $+1$, since every holistic observables(as example, $\left(I^{1} \otimes \sigma^{2}_{x} \right)$) appears twice, while QM predicts $-1$. This, obviously, is in contradiction with the QM eigenvalue relations given by Eqs.(3a)and (3b). 

For testing this contradiction, a statistically verifiable inequality has been proposed by Cabello\cite{cabellosi} is given by

\begin{equation}
\label{mc}
\left\langle\chi\right\rangle_{ncr} = \left\langle R_{1}\right\rangle +\left\langle R_{2}\right\rangle+\left\langle R_{3}\right\rangle + \left\langle C_{1}\right\rangle+\left\langle C_{2}\right\rangle-\left\langle C_{3}\right\rangle\leq 4
\end{equation}
Note that, for any state of two two-level system the quantum prediction $\left\langle\chi\right\rangle_{QM}=6$, thereby violating above inequality.

\section{A variant of Peres-Mermin proof} 

In order to provide a variant of Peres-Mermin proof\cite{mermin}, we use a new set of observables by replacing four out of nine holistic observables(in particular, those holistic observables which contains an identity operator) of the array given by Eq.(2) by the four new holistic observables. The modified Peres-Mermin array now looks very symmetric which is of the form  
\vskip -2.5cm
\begin{center}
\begin{equation}
\begin{array}{clrr}     
\sigma^{1}_{y}\otimes \sigma^{2}_{z} & \hskip 0.2cm \sigma^{1}_{z}\otimes \sigma^{2}_{y} &\hskip 0.3cm   {\sigma^{1}_{x}\otimes\sigma^{2}_{x}} \\
\\
\sigma^{1}_{z} \otimes \sigma^{2}_{x}  &\hskip 0.2cm    \sigma^{1}_{x}\otimes \sigma^{2}_{z}  &  \sigma^{1}_{y}\otimes\sigma^{2}_{y}\\
\\
\sigma^{1}_{x}\otimes\sigma^{2}_{y} &\hskip 0.2cm  \sigma^{1}_{y}\otimes\sigma^{2}_{x}& \sigma^{1}_{z}\otimes\sigma^{2}_{z}
\end{array}
\end{equation}
\end{center}

Now, for any quantum state of two two-level system the following quantum mechanical eigenvalue relations hold good:
\begin{subequations}
\begin{eqnarray}
\label{m1}
&&\hskip -0.8cmR^{\prime}_1\left|\Psi\right\rangle= R^{\prime}_{2}\left|\Psi\right\rangle= R^{\prime}_3\left|\Psi\right\rangle =  |\Psi\rangle\\
\nonumber
\\
\label{m2}
&&\hskip -0.8cmC^{\prime}_1\left|\Psi\right\rangle= C^{\prime}_2\left|\Psi\right\rangle=C^{\prime}_3\left|\Psi\right\rangle= -|\Psi\rangle
\end{eqnarray} 
\end{subequations}
where $R^{\prime}_{i=1,2,3}$ and $C^{\prime}_{i=1,2,3}$ are the rows and columns of the array given by Eq.(7) respectively.

In a given NCR model it is assumed that the individual measured values is fixed by the hidden variables follow the same operator relationship of QM. Then, from the QM eigenvalue relations given by Eqs.(\ref{m1}-\ref{m2}) one can write the following relations 
\begin{subequations}
\label{prime1}
\begin{eqnarray}
&&\hskip -0.8cm v(R^{\prime}_1)= v(R^{\prime}_{2})= v(R^{\prime}_3)= +1\\
\nonumber
\\
\label{prime2}
&&\hskip -0.8cm v(C^{\prime}_1)= v(C^{\prime}_2)=v(C^{\prime}_3)= -1
\end{eqnarray} 
\end{subequations}

Next, applying the product rule given by Eq.(1) it can be seen that the product of all $v(R^{\prime}_{i=1,2,3})$ and $v(C^{\prime}_{i=1,2,3})$ yields $+1$, since every holistic observables(as example, $\sigma^{1}_{x} \otimes \sigma^{2}_{x}$) appears twice, while QM predicts $-1$, contradicting the NCR models. 

It order to empirically test this incompatibility, we propose a new inequality that is different from the inequality given by Eq.(\ref{mc}) is as follows

\begin{equation}
\label{newmc}
\left\langle\delta\right\rangle_{ncr} = \left\langle R^{\prime}_{1}\right\rangle +\left\langle R^{\prime}_{2}\right\rangle+\left\langle R^{\prime}_{3}\right\rangle -\left\langle C^{\prime}_{1}\right\rangle-\left\langle C^{\prime}_{2}\right\rangle-\left\langle C^{\prime}_{3}\right\rangle\leq 4
\end{equation}
while  QM violates the above NCR inequality by predicting $\left\langle\delta\right\rangle_{QM}=6$ for any state of two two-state systems, such as, two spin-1/2 particles.  There are several such two two-state systems that have already been realised experimentally. For example, the path and the spin degrees of freedom of single neutrons that was used Hasegawa \emph{et. al.}\cite{hasegawa1} and a pair of trapped ions that is experimentally realised by Kirchmair \emph{et. al}\cite{nature}.  

\section{summary and conclusions}
In this work, we provide a suitable variant of Peres-Mermin proof for showing the contradiction between QM and the NCR models. For this purpose, we use a new set of holistic observables by replacing four out of nine holistic observables that were used in Peres-Mermin proof with four new holistic observables.  In order to empirically test such a  contradiction we proposed a new inequality that is violated by QM. Finally, we remark that  the incompatibility between QM and the NCR models demonstrated here can immediately be tested by the existing experimental techniques that have already been used\cite{nature,ams} for testing the NCR models.

\section*{Acknowledgments}
 Author is indebted to Dipankar Home and Guruprasad Kar for the insightful interactions in numerous occassions. This work is supported by the Research Associateship of Bose Institute, Kolkata. 
 
\end{document}